\documentclass[12pt,preprint]{aastex}

\usepackage{graphicx}
\usepackage{layout}
\usepackage[latin1]{inputenc}
\usepackage{amssymb}
\usepackage{natbib}

\slugcomment{To appear in Ap. J.}
\shorttitle{Nuclear to host relation of high redshift quasars}
\shortauthors{Kotilainen et al.}

\def\ref{\par\noindent\hangindent 20 pt}

\def\mincir{\ \raise -2.truept\hbox{\rlap{\hbox{$\sim$}}\raise5.truept %MC
\hbox{$<$}\ }}  %
\def\magcir{\ \raise -2.truept\hbox{\rlap{\hbox{$\sim$}}\raise5.truept %
\hbox{$>$}\ }}

\begin{document}

\title{The nuclear to host galaxy relation of high redshift quasars}

\author{Jari K. Kotilainen}
\affil{Tuorla Observatory, University of Turku, V\"ais\"al\"antie 20, 
FI--21500 Piikki\"o, Finland}
\email{jarkot@utu.fi}

\author{Renato Falomo}
\affil{INAF -- Osservatorio Astronomico di Padova, Vicolo dell'Osservatorio 5, 
35122 Padova, Italy}
\email{falomo@pd.astro.it}

\author{Marzia Labita and Aldo Treves}
\affil{Universit\`a dell'Insubria, via Valleggio 11, I-22100 Como, Italy}
\email{marzia.labita@uninsubria.it; treves@mib.infn.it}

\and

\author{Michela Uslenghi}
\affil{INAF-IASF Milano, Via E. Bassini 15, Milano I-20133, Italy}
\email{uslenghi@iasf-milano.inaf.it}

\begin{abstract}

We present near-infrared imaging obtained with ESO VLT+ISAAC of 
the host galaxies of a sample of low luminosity quasars in the redshift range 
1 $< z <$ 2, aimed at investigating the relationship between the nuclear and 
host galaxy luminosities at high redshift. This work complements our previous 
systematic study to trace the cosmological evolution of the host galaxies of 
high luminosity quasars (Falomo et al. 2004). The new sample includes 
15 low luminosity quasars, of which nine are radio--loud (RLQ) and six are 
radio--quiet (RQQ). They have similar distribution of redshift and 
optical luminosity, and together with the high luminosity quasars they cover 
a large range ($\sim$4 mag) of the quasar luminosity function. For all 
the observed objects, except one RQQ and one RLQ, we have been able to derive 
the global properties of the surrounding nebulosity. The host galaxies of 
both types of quasars are in the range of massive inactive ellipticals 
between $L^*$ and 10$L^*$. RLQ hosts are systematically more luminous than 
RQQ hosts by a factor of $\sim$2. This difference is similar to that found for 
the high luminosity quasars. This luminosity gap appears to be independent of 
the rest-frame $U$-band luminosity but clearly correlated with the rest-frame 
$R$-band luminosity.  
The color difference between the RQQs and the RLQs is likely 
a combination of an intrinsic difference in the strength of the thermal and 
nonthermal components in the SEDs of RLQs and RQQs, and a selection effect 
due to internal dust extinction.
For the combined set of quasars, we find 
a reasonable correlation between the nuclear and the host luminosities. 
This correlation is less apparent for RQQs than for RLQs. 
If the $R$-band luminosity is representative of 
the bolometric luminosity, and assuming that 
the host luminosity is proportional to the black hole mass, as observed in 
nearby massive spheroids, quasars emit with a relatively narrow range of power 
with respect to their Eddington luminosity and with the same distribution for 
RLQs and RQQs.

\end{abstract}


\section{Introduction}

Low redshift (z $\leq$ 0.5) quasars are predominantly hosted by luminous, 
massive, bulge-dominated galaxies (e.g. McLeod \& Rieke 1994; 
Taylor et al. 1996; Bahcall et al. 1997; Percival et al. 2000; 
Hamilton, Casertano \& Turnshek 2002; Dunlop et al. 2003; 
Pagani, Falomo \& Treves 2003; Floyd et al. 2004). More specifically, 
the nature of the host depends on nuclear luminosity in the sense that 
high luminosity quasars are exclusively hosted in elliptical galaxies, 
while fainter radio quiet quasars can also be found in early-type spirals 
(Hamilton et al. 2002; Dunlop et al. 2003). This trend is consistent with 
the fact that low luminosity type 1 AGN (e.g. Seyfert 1s) are only found in 
spiral galaxies (e.g. Kotilainen \& Ward 1994; Hunt et al. 1997). 

This scenario is consistent with the fact that practically all nearby 
massive spheroids (ellipticals and bulges of early-type spirals) have an 
inactive supermassive black hole (BH) in their centers (see e.g. Barth 2004; 
Ferrarese 2006 for recent reviews) and that more massive bulges host the most 
massive BHs. This suggests that episodic quasar activity (with a varying 
duty cycle) may be very common in massive galaxies and that the nuclear power 
depends on the mass of the galaxy. Powerful quasar activity is in fact only 
found in the most luminous (massive) galaxies (Hamilton et al. 2002; 
Falomo, Carangelo \& Treves 2003; Kauffmann et al. 2003).
At low redshift, the mass of the BH is correlated to the luminosity and 
the velocity dispersion of the bulge (e.g. Magorrian et al. 1998, 
Ferrarese \& Merritt 2000, Gebhardt et al. 2000, McLure \& Dunlop 2002, 
Marconi \& Hunt 2003, Bettoni et al. 2003, H\"aring \& Rix 2004). 
Furthermore, the strong cosmological evolution of the quasar population, 
with a co-moving space density peak at z $\sim$2--3 before rapidly declining 
to its low present value (Dunlop \& Peacock 1990; 
Warren, Hewett, \& Osmer 1994; Boyle 2001) is similar to the BH mass 
accretion rate and the evolution of the cosmic star formation history 
(Madau, Pozzetti, \& Dickinson 1998; Franceschini et al. 1999; 
Steidel et al. 1999; Chary \& Elbaz 2001, Barger et al. 2001, 
Yu \& Tremaine 2002, Marconi et al. 2004). Therefore, determining how 
the properties of the galaxies hosting quasars evolve with the cosmic time may 
be crucial to investigate the fundamental link between the formation and 
evolution of massive galaxy bulges and their nuclear activity, and to reveal 
whether BHs and spheroids really grow synchronously. 

The detection of the host galaxies for high redshift objects and 
the characterization of their properties is rather challenging because 
the host galaxy becomes rapidly very faint compared to the nucleus. In order 
to cope with this, imaging with high spatial resolution and S/N together with 
a well defined point spread function (PSF) for modeling the images are of 
crucial importance. 

In a previous work (Falomo et al.~2004), with the 8m 
Very Large Telescope (VLT) and ISAAC, we have carried out 
a systematic imaging study under 
excellent seeing conditions (median $\sim$0\farcs4 FWHM) of 17 quasars 
(10 radio-loud quasars [RLQ] and seven radio-quiet quasars [RQQ]) 
in the redshift range 1 $<$ z $<$ 2 to characterize their host galaxies. 
We found that the luminosity evolution of both RLQ and RQQ hosts until 
z $\sim$2 is  consistent with that of massive ellipticals undergoing 
passive evolution. There is no significant decrease in the host mass until 
z $\sim$2 as would be expected in the models of hierarchical formation of 
massive ellipticals (Kauffmann \& Haehnelt 2000). 
Note, however, that more recent hierarchical models including 
e.g. feedback due to AGN and supernovae (e.g. Granato et al. 2004; 
Bower et al. 2006) are in agreement with the existence of 
a substantial population of massive ellipticals out to at least z $\sim$2.
We also found evidence that 
RLQ hosts are systematically more luminous (massive) by a factor $\sim$2 
than RQQ hosts at all redshifts. A similar result was obtained by 
Kukula et al. (2001) using a smaller sample of quasars at z $\sim$0.9 and 
z $\sim$1.9. Little correlation was found between the nuclear and 
the host luminosities. Note that at low redshift some claims of a correlation 
between the two quantities have been reported in the literature 
(e.g. McLeod \& Rieke 1994; Bahcall et al. 1997; Hooper, Impey \& Foltz 1997), 
whereas some other studies of low redshift quasars (e.g. Dunlop et al. 2003) 
have not found a correlation for high luminosity quasars. 
Obviously, selection effects due to the difficulty of detecting faint galaxies 
hosting bright quasars, or vice versa, of detecting weak quasars located in 
bright host galaxies, may combine to form a spurious correlation 
(e.g. Hooper et al. 1997).

The majority of quasars studied at high redshift, including those in 
Falomo et al. (2004), belong to the bright end of quasar luminosity function, 
due to the selection effects in flux-limited samples, 
e.g. the redshift - luminosity degeneracy. In this paper, we present 
an imaging study of a sample of lower luminosity quasars (by $\sim$2 mag 
on average with respect to that of Falomo et al. 2004), to study 
the dependence of host properties on nuclear luminosity at high redshift. 
We have therefore a well matched sample of quasars and we are in 
a good position to study the significance of any correlation between 
the nuclear and the host luminosities. In section 2 we describe our  sample, 
in section 3 we report the observations and in section 4 we describe 
the data analysis. Our results for the observed quasars and their comparison 
with the host luminosities of quasars derived from other samples, 
the relationship between host and nuclear luminosities, and the cosmic 
evolution of RLQ and RQQ host galaxies are discussed in section 5. Summary of 
our results and directions for future work are given in Section 6. We adopt 
the concordance cosmology with 
H$_0$ = 70 km s$^{-1}$ Mpc$^{-1}$, $\Omega_m$ = 0.3 and $\Omega_\Lambda$ = 0.7 

\section{The sample}

The sample of low luminosity (hereafter LL) quasars was defined to be matched 
in redshift with the sample of high luminosity (hereafter HL) quasars of 
Falomo et al. (2004). It was extracted from the quasar catalogue of 
Veron-Cetty \& Veron (2003) requiring: 1.0 $<$ z $<$ 2.0, $M_V <$ -25.0 at 
z = 1.0, increasing to $M_V <$ -26.3 at z = 2.0, 
--60$^\circ$ $<\delta <$ -8$^\circ$, and having sufficiently bright stars 
within 1 arcmin of the quasar in order to allow a reliable characterization of 
the PSF. 
Our choice of a slightly redshift dependent magnitude limit 
guarantees an optimal matching in redshift between the HL and LL subsamples, 
while roughly corresponding to a simple magnitude limit at $M_V\sim -25.8$.
We included both RLQs and RQQs in order to investigate the difference 
between the host galaxies of the two types of quasars. Importantly, 
the LL and HL RLQ and RQQ subsamples are well matched in both their redshift 
and optical/blue luminosity distribution. This selection yielded in total 
20 LL quasars, of which 15 quasars were imaged, nine RLQs and six RQQs 
(Table 1). There is no statistically significant difference in the properties 
of the original and the observed samples. Fig. 1 shows the distribution of 
the observed quasars in the redshift--optical luminosity plane compared with 
those for HL quasars in Falomo et al. (2004) and with all the quasars in 
Veron-Cetty \& Veron (2003).

\section{Observations}

Deep images of the quasars in the $H$- or $K$-band were obtained using 
the near-infrared (NIR) ISAAC camera (Cuby et al. 2000), mounted on UT1 (Antu) 
of VLT at the European Southern Observatory (ESO) in Paranal, Chile. 
The Short Wavelength (SW) arm of ISAAC is equipped with a 1024 x 1024 px 
Hawaii Rockwell array, with a pixel scale of 0\farcs147 px$^{-1}$, 
giving a field of view of $\sim$150 x 150 arcsec. The observations were 
performed in service mode in the period between 2004 October and 2005 January. 
A detailed journal of observations is given in Table 1.  The seeing, 
as derived from the full width half maximum (FWHM) size of the image of stars 
in each frame, was consistently excellent during the observations, 
ranging from  $\sim$0\farcs3 to $\sim$0\farcs5 (average and median 
FWHM = 0\farcs4).
The choice of observing in the $H$- and $K$-band, for objects 
at below and above z = 1.4, respectively, was motivated by observing the same 
rest-frame wavelengths as a function of redshift.

Total integration times were 36 minutes per target. The images were secured 
using individual exposures of 2 minutes per frame, and a jitter procedure 
(Cuby et al. 2000), which produces a set of frames at randomly offset 
telescope positions within a box of 10 x 10 arcsec centered on 
the first pointing. Data reduction was performed by the ESO pipeline for 
jitter imaging data (Devillard 1999). Each frame was flat-fielded by 
a normalized flat field obtained by subtracting ON and OFF images of 
the illuminated dome, after interpolating over bad pixels. Sky subtraction was 
derived by median averaging sky frames from the 10 frames nearest in time. 
The reduced frames were aligned to sub-pixel accuracy using a fast 
object detection algorithm, and co-added after removing spurious pixel values. 
Photometric calibration was performed using standard stars observed during 
the same night. The estimated internal photometric accuracy is $\pm$0.03 mag.

\section{Two-dimensional data analysis}

Two-dimensional data analysis has been carried out using AIDA, 
Astronomical Image Decomposition and Analysis (Uslenghi \& Falomo, in prep.), 
a software package specifically designed to perform two-dimensional 
model fitting of quasar images, providing simultaneous decomposition into 
nuclear and host components. The analysis consists of two main parts: 
a) PSF modeling and b) quasar decomposition.

\subsection{PSF modeling}

To detect the host galaxies of quasars and to characterize their properties, 
the key factors are the nucleus-to-host magnitude ratio and the seeing 
(the shape of the PSF). The most critical part of the analysis is thus to 
perform a detailed PSF modeling for each frame. This is based on fitting 
a parameterized bidimensional model to the field stars, that are selected 
based on FWHM, roundness and signal-to-noise ratio. A sufficiently bright, 
saturated star was included in the list of reference stars in order to model 
the shape of the faint wing of the PSF, against which most of the signal from 
the surrounding nebulosity will be detected. The relatively large 
field of view of ISAAC ($\sim$2\farcm5) and the constraint on the quasar 
selection to have at least one bright star within 1 arcmin from the quasar, 
allowed us to reach this goal and thus to perform a reliable characterization 
of the PSF. Images with a large number of stars distributed over 
the field of view have been checked to account for any possible 
positional dependence of the PSF. No significant variations were found and in 
this analysis the PSF is assumed to be spatially invariant, 
i.e., the same model has been fitted simultaneously to all the reference stars 
of the image.

For each source, a mask was built to exclude contamination from 
nearby sources, bad pixels and other defects affecting the image. The local 
background was computed in a circular annulus centered on the source, and its 
uncertainty was estimated from the standard deviation of the values computed 
in sectors of concentric sub-annuli included in this area. The region to be 
used in the fit was selected by defining an internal and an external radius of 
a circular area. Setting the internal radius to a non-zero value allows 
excluding the core of bright, saturated stars.

\subsection{Quasar host characterization}

Once a suitable model of the PSF was determined, the quasar images were first 
fitted with only the PSF model in order to provide a first indication of 
a deviation from the PSF shape. 
Then the object was fitted with a point source plus a galaxy modeled as a 
de Vaucouleurs r$^{1/4}$ or a disk model convolved with the PSF for 
the host galaxy, adding a scaled PSF to represent the nucleus. 
If the residuals did not reveal any significant deviation, 
based on 
the comparison of the $\chi^{2}$ values between PSF-only and PSF+host models, 
the object was considered unresolved. 

With this procedure we can derive the luminosity and the scale-length of 
the host galaxies and the luminosity of the nuclei. An estimate of the errors 
associated with the computed parameters was obtained by simulating the process 
with synthetic data. Simulated quasar images were generated adding noise to 
the best fit model, then the fit procedure was applied to these images, 
producing a "best fit" combination of parameter values for each image. 
For each parameter, the standard deviation of the best-fit values gives 
an estimate of the uncertainty on the parameters. Obviously, this procedure 
does not take into account systematic errors generated by non-perfect modeling 
of the PSF, which can be roughly estimated by comparing results obtained with 
different PSF models, statistically consistent with the available data. 
In our worst case, Q 0335-3546, for example, this effect produces 
an uncertainty of $\sim$0.3 mag on the brightness of the host galaxy. 
Instead, in cases with a large number of suitable reference stars, 
the uncertainty is dominated by the noise. Upper limits to host magnitudes of 
unresolved objects were computed by adding a galaxy component to the PSF and 
varying its surface brightness until the model profile was no longer 
consistent with the observed profile.

While the total magnitude of the host galaxy can be derived with a typical 
internal error of 0.2 - 0.7 mag (0.4 mag on average), the scale-length is 
often poorly constrained. This depends on the degeneracy that occurs between 
the effective radius $r_e$ and the surface brightness $\mu_e$ 
(see Taylor et al.~1996). 

At high redshifts, it becomes difficult to distinguish between 
exponential disk and bulge models from the luminosity distributions. 
In this work, we have assumed that the host galaxies can be represented as 
elliptical galaxies following a de Vaucouleurs model. This is supported by 
the strong evidence at low redshift for the predominance of bulge dominated 
hosts of quasars (e.g. Hamilton et al. 2002; Dunlop et al. 2003; 
Pagani et al. 2003 and references therein). 
Table 2 shows that for practically all the RLQs in our sample, 
we formally find a better fit (a lower $\chi^{2}$ value) using 
a de Vaucouleurs model. For the RQQs, the situation is reverse, 
with 4/5 objects formally having a better fit with a disk model. In most of 
these cases, however, the difference in the $\chi^{2}$ value between 
the models is negligible.
Note that adopting a disk model would result in fainter host galaxies, 
by $\sim$0.5 mag on average, but this would not introduce systematic 
differences that would affect our conclusions.

\section{Results}

In Fig. 2, we report for each observed quasar the image of the quasar, 
the best-fitting host galaxy model after subtracting a scaled PSF, 
the residuals after fitting the model, the radial brightness profile and 
the best fit using the procedure described above. The parameters of 
the best fit, together with their estimated uncertainty, are given in Table 2. 
All quasars except two (RQQ Q 1045+056 and RLQ PKS 0805-07) are resolved. 
This is quantified in Table 2 by comparing the reduced $\chi^2_\nu$ value of 
the best fit including a host galaxy model with that obtained from 
the best fit performed only with the PSF model. 

In Table 3, we report the absolute magnitudes and the effective radii for each 
quasar host. As our observations in the redshift range $1<z<2$ were obtained 
in the $H$ and $K$ filters, the detections of the host galaxies roughly 
correspond to rest-frame $7000-8000$ \AA{}. This is also the case for 
the HL sample of Falomo et al. (2004). On the other hand, Kukula et al. (2001) 
and Ridgway et al. (2001) observed $z\sim1.9$ quasars with the HST 
filter F165M, which corresponds to rest-frame $5500-6000$ \AA. 
Therefore, in order to refer all these observations to the same band 
(and to minimize the color and K-corrections), we transformed 
observed magnitudes into absolute magnitudes in the $R$-band. 
Moreover, the use of $R$-band magnitudes offers the possibility of 
a relatively easy comparison with the majority of the published low redshift 
quasar host studies. To perform the color and K-correction transformations, 
we assumed an elliptical galaxy template (Mannucci et al. 2001) 
for the host galaxy magnitudes, and a composite quasar spectrum 
(Francis et al. 1991) for the nuclear magnitudes, 
Note that the K-correction in the observed K-band is almost independent 
of galaxy type up to z $\sim$2, whereas in the observed H-band at 
1 $<$ z $<$ 1.5, the K-correction depends on the assumed host galaxy template, 
being $\sim$0.1 - 0.25 mag larger for elliptical galaxies than for 
spiral (Sc) galaxies (Mannucci et al. 2001).

\subsection{Properties of the host galaxies of quasars at 1 $<$ z $<$ 2}

In Fig. 3 we compare the absolute $R$-band magnitudes of the quasar 
host galaxies versus redshift in this work, with the HL quasar hosts 
(Falomo et al. 2004) and with quasar hosts from Kukula et al. (2001) 
and Ridgway et al. (2001). Note that the z $\sim$2 sample of 
Kukula et al. (2001) has on average similar nuclear luminosity 
($M_R$ = --24.9 $\pm$ 0.9) to our LL sample ($M_R$ = --24.9 $\pm$ 1.0), 
whereas the RQQs in Ridgway et al. (2001) at z $\sim$1.8 are 
significantly fainter ($M_R$ = --23.3 $\pm$1.2). In order to treat these 
literature data homogeneously, we have considered the published 
apparent magnitudes in the $H$ and $K$-bands and transformed them to $M_R$ 
following our procedure (K-correction, cosmology and color correction). 
The average absolute $R$-band magnitudes of the host galaxies of the samples 
of LL RLQs and RQQs (this work) and HL RLQs and RQQs (Falomo et al. 2004) 
are given in Table 4, column 6. 

Almost all the observed quasars have host galaxies with luminosity ranging 
between $L^*$ and $10 L^*$, where $M^*$(R)$\sim-21.2$ (Gardner et al. 1997; 
Nakamura et al. 2003) is the characteristic luminosity of the Schechter 
luminosity function for elliptical galaxies. 

For the LL quasars, 
there is a systematic difference in the luminosity between RLQ and RQQ 
host galaxies of $\sim$0.9 mag. Similar difference has been found in 
many previous studies for quasars at low redshift (Bahcall et al. 1997; 
Hamilton et al. 2002; Dunlop et al. 2003) and high redshift 
(Kukula et al. 2001). Floyd et al. (2004) found no difference 
between the RLQ and RQQ host luminosities at z $\sim$0.4 
($<M_V>$(RQQ)=--23.35, $<M_V>$(RLQ)=--23.07 for elliptical host galaxies), 
but we note that their subsamples are not well matched in 
nuclear luminosity.
The difference found in this work is also similar to that found for HL quasars 
($\sim$0.7 mag; Falomo et al. 2004). and our new results thus confirm this 
offset, based on a larger statistical sample and a larger luminosity interval.

\subsection{The relation between nuclear and host luminosities}

If the mass of the central BH is proportional to the mass and thus to 
the luminosity of the spheroid of the host galaxy, as it is observed for 
nearby inactive early-type galaxies (Kormendy \& Richstone 1995; 
Magorrian et al. 1998), and if the quasar emits at a roughly fixed fraction of 
the Eddington luminosity, one would expect a correlation between 
the luminosity of the nucleus and that of the host galaxy. 
However, nuclear obscuration, beaming, and/or an intrinsic spread in 
the accretion rate and accretion-to-luminosity conversion efficiency, 
could destroy this correlation.

Our combined sample of LL and HL quasars in this work and in 
Falomo et al. (2004) is designed to explore a large range of 
nuclear luminosity (-23.5 $< M_V <$ -28; H$_0$ = 50 km s$^{-1}$ Mpc$^{-1}$, 
q$_0$ = 0, corresponding to -22.8 $< M_V <$ -27.1 in our adopted cosmology) 
and can therefore be used to investigate this issue. In Table 4 
we report the average values of the rest frame $U$-band absolute magnitudes 
for the four subsamples (HL RLQ, LL RLQ, HL RQQ and LL RQQ). These values are 
derived from the $V$-band apparent magnitudes reported in 
Veron--Cetty \& Veron (2003), K-corrected and color corrected following 
the procedure described above. A correction for the Galactic extinction was 
applied following Schlegel, Finkbeiner \& Davis (1998). 

Both in the LL and HL samples, the nuclear $U$-band luminosities of the RLQs 
and RQQs are matched within 0.1~mag. On the other hand, 
considering the rest-frame $R$-band nuclear luminosities, RLQs appear 
more luminous than RQQs by $\sim1\,$mag. Note that also in the sample of 
Kukula et al. (2001), the quasars are well matched in the $U$-band, but not in 
the $R$-band, where again the RLQs appear more luminous than the RQQs. 
These results, therefore, suggest that (at least in the redshift range 
considered here) there is a systematic color difference between the nuclei of 
RLQs and those of RQQs, in the sense that RLQs are redder than RQQs by 
$\sim0.8\,$mag in the rest-frame $U-R$ color. 
Indeed, there is no apparent difference 
between the UV-to-NIR spectral properties of RLQs and RQQs in the well known 
average quasar spectral energy distribution (SED) of Elvis et al.~(1994), 
but the considered sample is biased toward X-ray and optically bright 
(i.e. bluer) quasars. Some hint of a possible difference between the SED of 
RLQs and RQQs was reported by Barkhouse \& Hall (2001), who observed 
a greater  NIR-to-optical luminosity ratio of RLQs with respect to RQQs in 
a large sample of quasars detected by 2MASS. 
Furthermore, Francis et al.~(2000) found that the optical--NIR continuum 
is significantly redder in radio selected RLQs from 
the PKS Half-Jansky Flat-Spectrum survey than in optically selected RQQs from 
the Large Bright Quasar Survey. 

This effect may be interpreted as due to a differential extinction by dust or 
to an intrinsic difference of the strength of thermal and non thermal emission 
components in the SEDs of RQQs and RLQs. For instance, in the case of 
flat spectrum quasars, one could expect to observe an enhanced non-thermal 
(synchrotron) component which contaminates the SED more in the near-IR than in 
the $UV$. Of course, this would suggest that the near-IR luminosity is not 
a good tracer of the bolometric emission. However, Francis et al. (2000) 
find that this effect is not sufficient to describe the spectral shape of all 
the sources in their radio-selected sample: about the 50 per cent of their 
PKS QSOs are more likely to be reddened by dust. We believe that, 
with present data, both explanations (synchrotron contamination and 
dust extinction) are viable; however, in the specific case of our sample 
of RLQs, the hypothesis of synchrotron contamination is weakened because 
one third of the objects are steep spectrum radio sources (viewed further away 
from the jet axis than flat spectrum radio quasars), and there is no 
correlation between the radio spectral index and 
the $U-R$ (observed $V-K$) color. If indeed extinction by dust is 
the dominant effect, then the $R$-band would be a better tracer of 
the bolometric luminosity than the $U$-band. Moreover, we note that in 
the rest-frame $U$, $B$ region, the SED of a QSO is contaminated by 
the variable thermal emission in the accretion disk (the big blue bump), 
suggesting again that the $R$-band luminosity is a better indicator of 
the total nuclear emission.

In Fig. 4, we compare the rest-frame $R$-band host and nuclear luminosities of 
the HL and LL quasars, together with quasars from Kukula et al. (2001) and 
Ridgway et al. (2001), both for individual quasars (upper panel) and for 
the samples (lower panel). The resulting Spearman rank 
correlation coefficients (R$_S$), and the probabilities of obtaining 
the observed R$_S$ values if no correlation is present are given in Table 5. 
Although this comparison is based on non-complete samples and is subject 
to uncertainties due to small number statistics, we find some correlation 
for the full sample of RLQs and RQQs (R$_S$=0.49, and the probability of 
no correlation, $P$(nc) $\sim10^{-3}$). This correlation becomes modest 
for RLQs (R$_S$=0.36, $P$(nc)=0.1) and disappears altogether for RQQs 
(R$_S$=0.25, $P$(nc)=0.3). Generally, no such correlation has been found 
at low redshift (e.g. Dunlop et al. 2003; Pagani et al. 2003). 
However, interestingly, a similar trend to that found in this work is apparent 
considering the data given in Hamilton et al. (2002), who studied 
a large sample of $z < 0.46$ quasars. We derived luminosities from their 
reported nuclear and host galaxy apparent magnitudes, following our procedure, 
and we found a clear correlation for their full sample of quasars (R$_S$=0.56, 
$P$(nc)=$10^{-3}$) and for their RLQs (R$_S$=0.51, 
$P$(nc)=$10^{-2}$), while only a modest correlation is evident for 
their RQQs (R$_S$=0.38, $P$(nc)=0.1). This is an indication that 
the different trends of the nucleus--host luminosity relation displayed by 
RLQs and RQQs may be independent of redshift.

Assuming that the correlation between the central BH mass and the host galaxy 
luminosity holds up to $z\sim2$ and that the observed nuclear power is 
proportional to the bolometric luminosity, the observed 
nucleus -- host luminosity correlation can be interpreted as the result of 
an intrinsically narrow distribution of the Eddington ratio.
The observed scatter is then enhanced by the dispersion in 
the bulge luminosity -- BH mass correlation and by intrinsic differences in 
the accretion rates. This is consistent with the relationship between 
the host galaxy and maximum nuclear luminosity observed at lower redshift 
(e.g. Floyd et al.~2004).

\section{Summary and conclusions}

We have presented homogeneous high resolution  NIR images for a sample of 
15 low luminosity quasars in the redshift range 1 $<$ z $<$ 2, to characterize 
the properties and the cosmological evolution of their host galaxies, and to 
make a reliable comparison between RLQ and RQQ hosts. Together with 
the high luminosity quasars previously studied by us, they cover a large range 
($\sim$4 mag) of the quasar luminosity function. 

The quasar host galaxies follow the trend in luminosity of massive 
inactive ellipticals (between $L^*$ and 10$L^*$) undergoing simple 
passive evolution. However, RLQ hosts appear systematically more luminous 
(massive) than RQQ hosts by a factor of $\sim$2. This difference is similar to 
that found for the high luminosity quasars and our new observations indicate 
that this gap is apparently independent of the nuclear luminosity in 
the observed $V$-band (rest frame $U$-band) in the sense that at a fixed 
observed $B$-band luminosity, the host galaxies of RLQs are a factor of 
$\sim$2 brighter in the observed $H/K$-band (rest frame $R$-band) than those 
of RQQs.

However, if the $R$-band nuclear luminosity is considered, the gap in 
the host luminosity could be ascribed to a difference in the total 
nuclear power. In fact (see Fig. 4), the magnitude gap of the host luminosity 
corresponds to a similar gap in the nuclear $R$-band luminosity, 
suggesting that at a fixed host mass (and BH mass), the same bolometric power 
is emitted. 

For the combined sample of RQQs and RLQs, we find some correlation between 
the nuclear and the host luminosities, albeit with a large scatter, 
possibly due to a varying accretion efficiency. If the host luminosity is 
proportional to the black hole mass, quasars emit in a narrow range of power 
with respect to their Eddington luminosity. This range does not depend on 
redshift or on the radio properties of the quasars.

Determining the quasar host properties at even higher redshift, 
around the peak epoch of quasar activity (z $\sim$2.5) and beyond, 
requires very high S/N observations with a very narrow reliable PSF. We have 
an ongoing program to tackle this problem using NIR adaptive optics imaging 
with NACO on VLT for high luminosity quasars (Falomo et al. 2005,2007), 
and NIR non-adaptive optics with ISAAC on VLT for low luminosity quasars 
(Kotilainen et al., in prep.). Color information for the hosts 
(e.g. deep $R$-band imaging to target rest-frame UV emission), spectroscopy to 
estimate the BH masses of high redshift quasars, and the study of environments 
as a function of redshift and radio power, will also be addressed in 
future work. 

\acknowledgments

This work was partially supported by the Italian Ministry for University 
and Research (MIUR) under COFIN 2002/27145, ASI-IR 115 and ASI-IR 35, 
ASI-IR 73 and by the Academy of Finland (projects 8201017 and 8107775). 
This publication makes use of data products from 
the Two Micron All Sky Survey, which is a joint project of 
the University of Massachusetts and 
the Infrared Processing and Analysis Center/ 
California Institute of Technology, funded by 
the National Aeronautics and Space Administration and 
the National Science Foundation. This research has made use of 
the NASA/IPAC Extragalactic Database {\em(NED)} which is operated by 
the Jet Propulsion Laboratory, California Institute of Technology, 
under contract with the National Aeronautics and Space Administration.

\clearpage

% Figure 1      % z-M(V) distribution
%
\begin{figure}
\epsscale{1}
\plotone{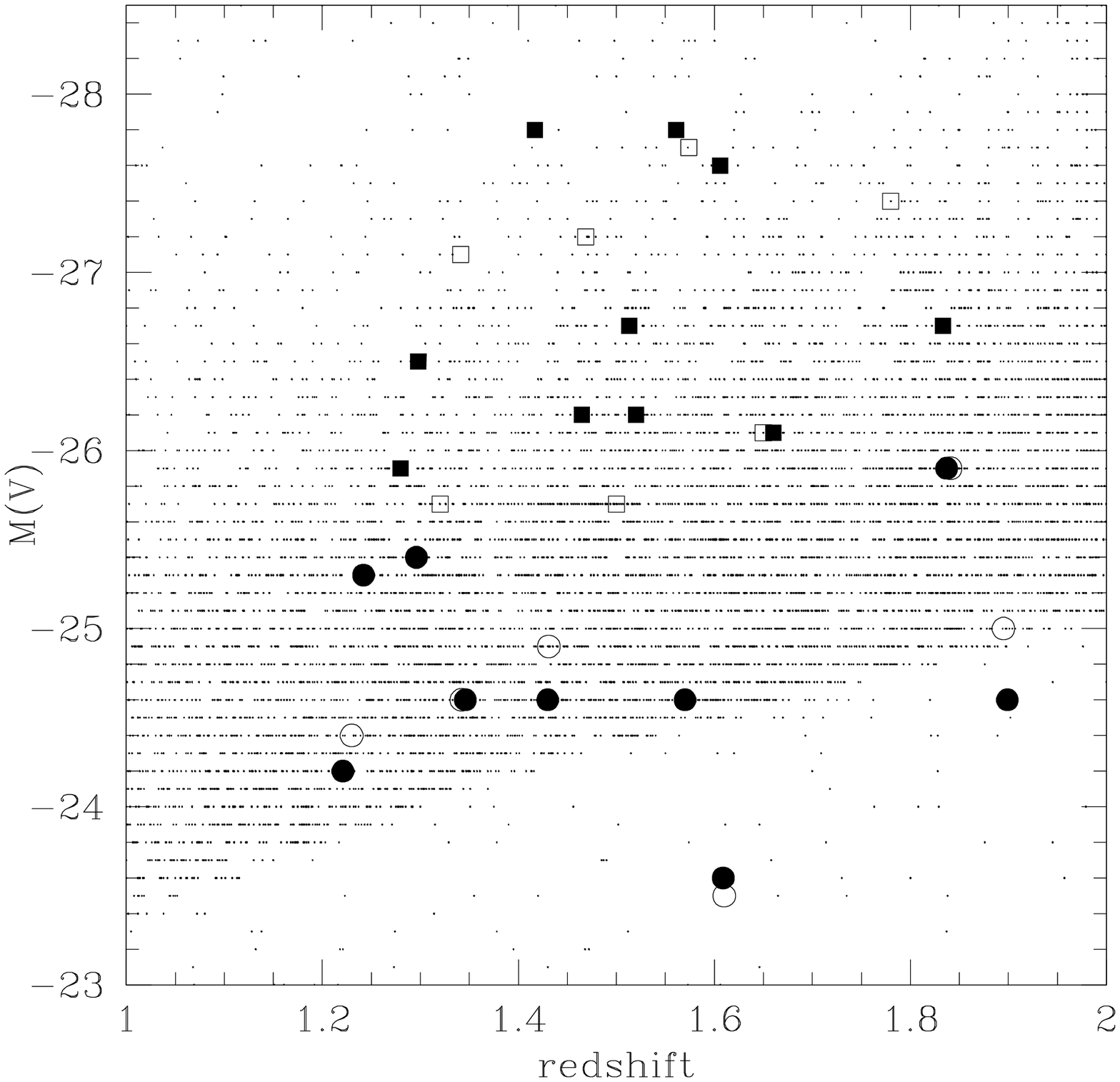}
%\centering
%\includegraphics[width=12cm]{f1.eps}
\caption{
The distribution of the low luminosity RLQs (filled circles) and RQQs 
(open circles) in the z - M$_V$ plane, compared with the high luminosity RLQs 
(filled squares) and RQQs (open squares) from Falomo et al. (2004) and, 
for reference, with all quasars at 1 $<$ z $<$ 2 in 
Veron-Cetty \& Veron (2003; small dots). 
\label{fig:MBz}}
\end{figure}

% Figure 2  % radial profile and fit
%
\begin{figure*}
\label{prof}
\epsscale{0.9}
%\plotone{f2a_new.eps}
\caption{Images of the central $\sim$7 x 7 arcsec region surrounding 
the quasars, from top to bottom, (a) the original image, (b) the image after 
subtracting a scaled PSF model (= the host galaxy) and (c) the residuals. 
These panels are on a linear scale from -3 $\sigma$ to +3 $\sigma$, 
where $\sigma$ is calculated from the sky noise. 
No interpolation, filtering or smoothing is applied. 
%Slight smoothing has been applied in panel (b) to show the host galaxies 
%more clearly.
Panel (d) shows the observed radial brightness profiles of the quasars 
(filled squares), superimposed to the fitted model consisting of the PSF 
(dotted line) and an elliptical (de Vaucouleurs law) galaxy convolved with 
its PSF (dashed line).  The solid line shows the composite model fit.}
\end{figure*}
\addtocounter{figure}{-1}%

\begin{figure*}
\label{prof1}
\epsscale{0.9}
%\plotone{f2b_new.eps}
\caption{continued.}
\end{figure*}
\addtocounter{figure}{-1}%
\begin{figure*}
\label{prof}
\epsscale{0.9}
%\plotone{f2c_new.eps}
\caption{continued.}
\end{figure*}
\addtocounter{figure}{-1}%
\begin{figure*}
\label{prof}
\epsscale{0.9}
%\plotone{f2d_new.eps}
\caption{continued.}
\end{figure*}
\addtocounter{figure}{-1}%
\begin{figure*}
\label{prof}
\epsscale{0.9}
%\plotone{f2e_new.eps}
\caption{continued.}
\end{figure*}

% Figure 3
\begin{figure}
\label{absmag}
\plotone{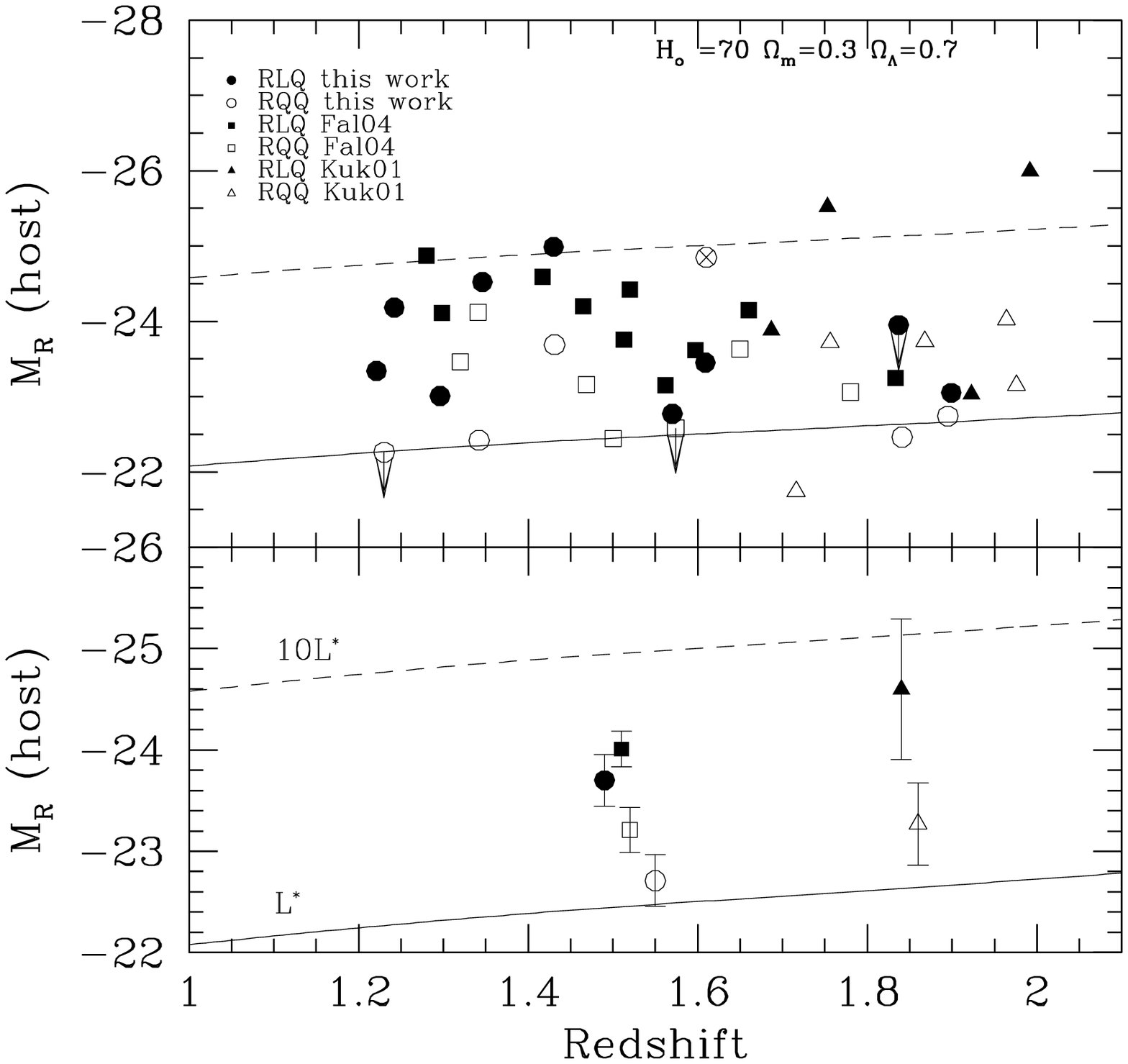}
\caption{ The $R$-band absolute magnitude of the quasar host galaxies 
versus redshift.  
The lines represent the expected behavior of a massive elliptical 
(at L$^*$ and 10L$^*$; {\it solid and dashed} lines ) 
undergoing simple passive evolution (Bressan, Chiosi \& Fagotto 1994). 
For symbols, see Fig. 1. 
Also included are the RLQs (filled triangles) and RQQs (open triangles) 
from Kukula et al. (2001). 
The arrows represent the upper limits of the host luminosity for 
the unresolved objects Q 1045+056, PHS 0805-07 (this work) and HE0935-1001 
(Falomo et al. 2004).
The object marked with a cross is TOL 1033.1-27.3, for which our results 
indicate that a disc-galaxy model is a better fit to the data than 
a de Vaucouleurs law. Moreover, this object has by far the lowest N/H ratio in 
the sample and may in fact not host a quasar.
The upper and lower panels show the data for individual quasars and for 
the sample averages, respectively.
}
\end{figure}

% Figure 4
\begin{figure}
\label{nuchost}
\plotone{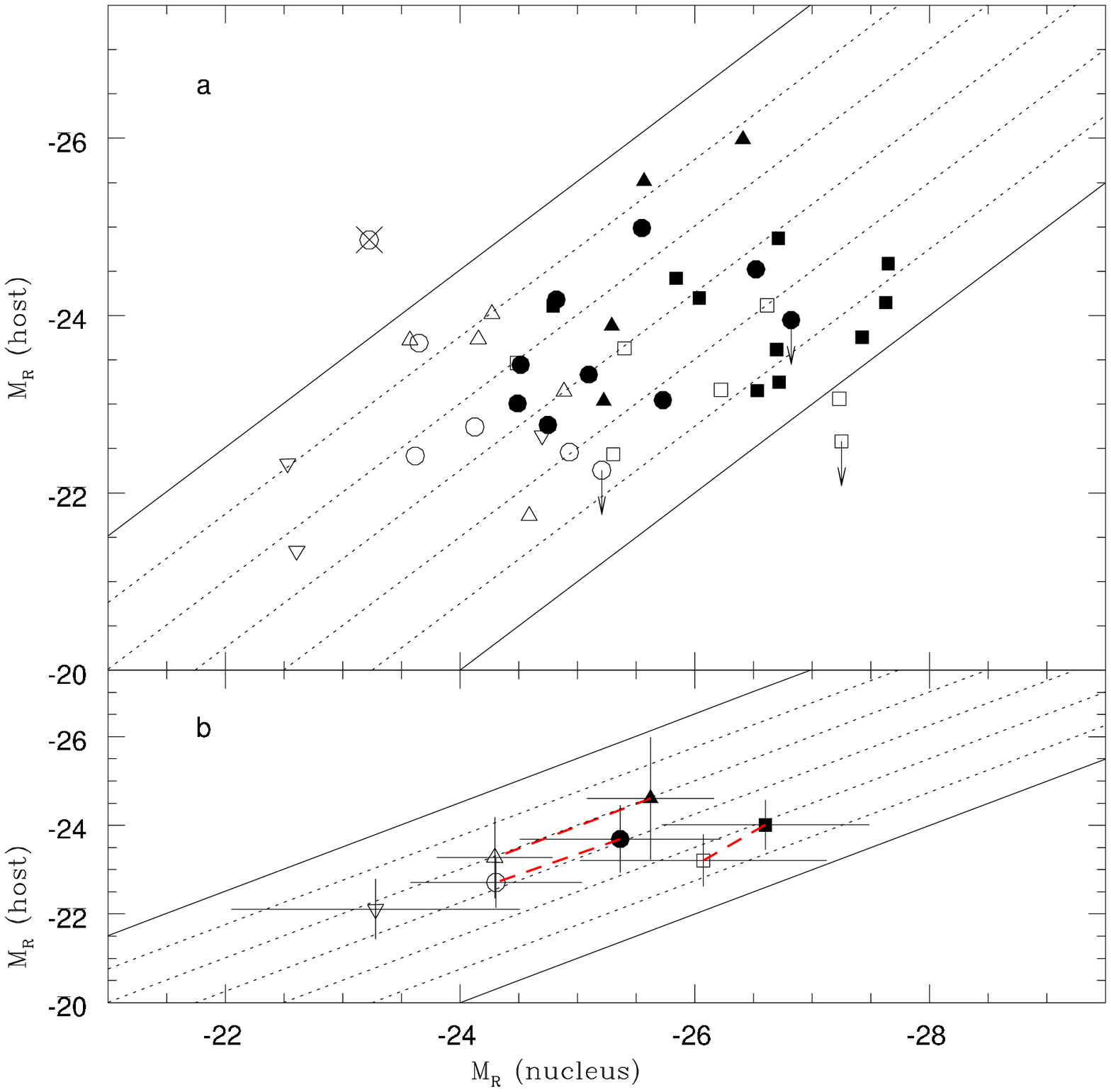}
\caption{
{\bf Upper panel:} 
The absolute magnitude of the nucleus compared with that of the host galaxy. 
For symbols, see Figs. 1 and 3. Open inverted triangles are the RQQs from
Ridgway et al. (2001). The arrows represent the upper limits of 
the host luminosity for the unresolved quasars Q 1045+056 and PKS 0805-07 
(this work) and HE0935-1001 (Falomo et al. 2004).
The object marked with a cross is TOL 1033.1-27.3 (see the caption of Fig.~3).
The diagonal lines represent the loci of constant ratio between host and 
nuclear emission. These can be translated into Eddington ratios assuming 
that the central BH mass - galaxy luminosity correlation holds up to 
z $\sim$ 2 and that the observed nuclear power is proportional to 
the bolometric emission. Separations between dotted lines correspond to a 
difference by a factor of 2 in the nucleus-to-host luminosity ratio. 
The two solid lines encompass a spread of 1.8dex in this ratio.
{\bf Lower panel:} the average values for the seven subsamples 
considered here (see Table 4), excluding TOL 1033.1-27.3. 
Note that, for each sample, the transition from 
RQQs to RLQs occurs at a roughly fixed fraction of the Eddington luminosity.
}
\end{figure}

%  TABLE 1
% Journal of observations
%
\begin{deluxetable}{l l l l l l l l}
\tablecolumns{8} \tablewidth{0pc} \tablecaption {TABLE 1}
\tablecaption  {Journal of the observations} \tablehead{
\colhead{Quasar} & \colhead{z} & \colhead{V\tablenotemark{a}}&
\colhead{Date} & \colhead{Filter} &
\colhead{$ T_{exp}$\tablenotemark{b}} &
\colhead{Seeing\tablenotemark{c}} & 
\colhead{N\tablenotemark{d}} \\
  &  &   &   & & (min) & (arcsec)  & }
\startdata \hline
\multicolumn{8}{c}{Radio Quiet Quasars}\\
\hline
Q 0335-3546      & 1.841 & 19.8 &  01/Nov/04 & K & 36 & 0.34 &  4\\
MS 0824.2+0327   & 1.431 & 20.2 &  24/Dec/04 & K & 36 & 0.33 &  5\\
2QZ J101733-0049 & 1.342 & 20.4 &  25/Dec/04 & H & 36 & 0.47 &  4\\
2QZ J101733-0203 & 1.895\tablenotemark{*} & 20.8 &  26/Dec/04 & K & 36 & 0.32 &  6\\
TOL 1033.1-27.3  & 1.610 & 21.8 &  26/Dec/04 & K & 36 & 0.32 &  4\\
Q 1045+056       & 1.230 & 20.3 &  26/Jan/05 & H & 36 & 0.52 &  3\\
\hline
\multicolumn{8}{c}{Radio Loud Quasars}\\
\hline
PKS 0258+011     & 1.221 & 20.5 &  01/Nov/04 & H & 36 & 0.51 &  3\\
PKS 0432-148     & 1.899 & 21.2 &  31/Oct/04 & K & 36 & 0.47 &  6\\
PKS 0442+02      & 1.430 & 20.5 &  01/Nov/04 & K & 36 & 0.37 &  7\\
PKS 0511-220     & 1.296 & 19.5 &  27/Nov/04 & H & 36 & 0.44 &  5\\
PKS 0805-07      & 1.837 & 19.8 &  25/Dec/04 & K & 36 & 0.37 &  7\\
PKS 0837+035     & 1.570 & 20.7 &  24/Dec/04 & K & 36 & 0.37 &  4\\
PKS 0845-051     & 1.242 & 19.4 &  21/Dec/04 & H & 36 & 0.38 &  5\\
PKS 1015-31      & 1.346 & 20.4 &  26/Dec/04 & H & 36 & 0.27 &  5\\
PKS 1046-222     & 1.609 & 21.7 &  25/Jan/05 & K & 36 & 0.37 &  3\\
\enddata
\tablenotetext{a} {Quasar $V$-band apparent magnitudes from
Veron-Cetty \& Veron (2003).} 
\tablenotetext{b} {Frame exposure time in minutes} \tablenotetext{c}
{The average FWHM in arcsec of all stars in the frame.}
\tablenotetext{d}
{Number of stars used for the PSF modelling of each field.}
\tablenotetext{*}
{The 2QZ catalogue contains two different redshift determinations for 
this object: $z_1=1.895$, $z_2=1.343$. 
We adopt $z_1=1.895$ here because 
the 2QZ spectrum from which this value is inferred shows stronger 
emission lines.
}
\end{deluxetable}

%  TABLE 2
% Results from fit of radial profiles 

\begin{deluxetable}{l l c l l l c c r r }
\tablecolumns{10} \tablewidth{0pc} \tablecaption {TABLE 2}
\tablecaption  {Results of the radial profile modeling.
} 
\tablehead{ \colhead{Quasar} &
\colhead{z}   & \colhead{Filter} &
\colhead{$m_{nuc}$\tablenotemark{a}}  &
\colhead{$m_{host}$\tablenotemark{a}} &  \colhead{r$_e$} &
\colhead{$\chi^{2}_{dV}$\tablenotemark{b}} &
\colhead{$\chi^{2}_{exp}$\tablenotemark{c}} &
\colhead{$\chi^{2}_{PSF}$\tablenotemark{d}}&
\colhead{DF\tablenotemark{e}}\\
  &  &   &   &  & (arcsec) & & & &}
\startdata \hline
\multicolumn{10}{c}{Radio Quiet Quasars}\\
\hline
Q 0335-3546      & 1.841 & K & 17.9 & 20.1$\pm$0.4 & 1.1$\pm$0.5 & 0.6 & 0.4 & 1.4 &  8 \\
MS 0824.2+0327   & 1.431 & K & 18.4 & 18.1$\pm$0.1 & 0.5$\pm$0.1 & 1.0 & 1.1 & 19.0&  15\\
2QZ J101733-0049 & 1.342 & H & 19.0 & 20.2$\pm$0.3 & 0.9$\pm$0.2 & 1.3 & 1.2 & 4.2 & 12 \\
2QZ J101733-0203 & 1.895 & K & 18.8 & 19.9$\pm$0.2 & 1.0$\pm$0.6 & 1.6 & 1.3 & 4.0 & 13 \\
TOL 1033.1-27.3  & 1.610 & K & 19.2 & 17.3$\pm$0.1 & 0.6$\pm$0.1 & 3.6 & 1.5 & 67.1&  35\\
Q 1045+056       & 1.230 & H & 17.4 & $>$20.1 & ... & - & - & 0.9                  & 11 \\
\hline
\multicolumn{10}{c}{Radio Loud Quasars}\\
\hline
PKS 0258+011     & 1.221 & H & 17.5 & 19.0$\pm$0.2 & 0.7$\pm$0.2 & 0.8 & 1.0 & 6.6 &  23\\
PKS 0432-148     & 1.899 & K & 17.2 & 19.6$\pm$0.3 & 1.2$\pm$0.2 & 1.2 & 1.5 & 2.0 &  17\\
-PKS 0442+02      & 1.430 & K & 16.5 & 16.8$\pm$0.1 & 0.8$\pm$0.2 & 1.1 & 3.5 &33.5&  39\\
PKS 0511-220     & 1.296 & H & 18.1 & 19.5$\pm$0.3 & 0.7$\pm$0.2 & 1.5 & 1.6 & 3.5 & 13 \\
PKS 0805-07      & 1.837 & K & 16.0 & $>$18.6 & ... & - & - & 1.6                  & 16 \\
PKS 0837+035     & 1.570 & K & 17.6 & 19.3$\pm$0.3 & 0.9$\pm$0.3 & 1.0 & 2.4 & 4.1 & 10 \\
PKS 0845-051     & 1.242 & H & 17.8 & 18.2$\pm$0.1 & 0.6$\pm$0.2 & 0.5 & 7.9 & 38.7&  9 \\
PKS 1015-31      & 1.346 & H & 16.1 & 18.1$\pm$0.1 & 0.9$\pm$0.2 & 1.0 & 2.3 & 17.7&  26\\
PKS 1046-222     & 1.609 & K & 17.9 & 18.7$\pm$0.2 & 0.7$\pm$0.2 & 1.2 & 1.2 & 10.3& 15 \\
\enddata
\tablenotetext{a} {Apparent magnitudes correspond to the indicated filter.}
\tablenotetext{b} {The reduced $\chi^2$ value of the fit with PSF and an 
elliptical host galaxy model.}
\tablenotetext{c} {The reduced $\chi^2$ value of the fit with PSF and an 
exponential disk host galaxy model.}
\tablenotetext{d} {The reduced $\chi^2$ value of the fit with only 
the PSF model. In the cases of Q 1045+056  and PKS 0805-07 the $\chi^2$ does not 
significantly improve when adding the galaxy component, therefore these
objects are indicated as unresolved.} 
\tablenotetext{e} {Number of degrees of freedom.} 
\end{deluxetable}

%  TABLE 3
% Properties of Quasars  and Host galaxies

\begin{deluxetable}{l l c c c c c l}
\tablecolumns{8}
\tablewidth{0pc}
\tablecaption {TABLE 3}
\tablecaption  {Properties of the quasars and their host galaxies.}
\tablehead{
\colhead{Quasar} &  \colhead{z}   &
\colhead{$\mu_e$\tablenotemark{a}} &  
\colhead{$M_R(nucl)$\tablenotemark{b}} & \colhead{$M_R(host)$\tablenotemark{b}}  
& \colhead{N/H\tablenotemark{c}} & \colhead{Re}  
& \colhead{$M_U(tot)$\tablenotemark{d}}\\
  &  &   &   &     & &(kpc)& }
\startdata
\hline
\multicolumn{8}{c}{Radio-quiet quasars}\\
\hline
Q 0335-3546      & 1.841 & 15.5 &  -24.9 & -22.5    &     9.12 &  9.5	 &  -26.0  \\
MS 0824.2+0327	 & 1.431 & 11.8 &  -23.7 & -23.7    &     1.00 &  4.4	 &  -25.1  \\
2QZ J101733-0049 & 1.342 & 15.0 &  -23.6 & -22.4    &     3.02 &  7.0	 &  -24.7  \\
2QZ J101733-0203 & 1.895 & 14.9 &  -24.1 & -22.7    &     3.63 &  8.1	 &  -25.2  \\
TOL 1033.1-27.3  & 1.610 & 11.2 &  -23.2 & -24.9    &     0.21 &  4.9	 &  -23.9  \\
Q 1045+056 	 & 1.230 &  -	&  -25.2 & $>$-22.3 & $>$14.4 &   - &  -24.5  \\
\hline
\multicolumn{8}{c}{Radio-loud quasars}\\		        	
\hline							        	
PKS 0258+011	 & 1.221 & 13.1 &  -25.1 & -23.3    &     5.25 &  5.3	 &  -24.5  \\
PKS 0432-148     & 1.899 & 15.0 &  -25.7 & -23.1    &    11.0 &  9.7	 &  -26.3  \\
PKS 0442+02      & 1.430 & 11.5 &  -25.6 & -25.0    &     1.74 &  6.9	 &  -25.2  \\
PKS 0511-220	 & 1.296 & 13.7 &  -24.5 & -23.0    &     3.98 &  5.5	 &  -25.5  \\
PKS 0805-07      & 1.837 &  -	&  -26.8 & $>$-24.0 & $>$13.2 &   - &  -26.5  \\
PKS 0837+035     & 1.570 & 14.3 &  -24.8 & -22.8    &     6.31 &  7.8	 &  -24.8  \\
PKS 0845-051     & 1.242 & 12.2 &  -24.8 & -24.2    &     1.74 &  5.0	 &  -25.4  \\
PKS 1015-31      & 1.346 & 12.8 &  -26.5 & -24.5    &     6.31 &  7.2	 &  -24.9  \\
PKS 1046-222     & 1.609 & 13.1 &  -24.5 & -23.5    &     2.51 &  6.2	 &  -23.9  \\
\enddata
\tablenotetext{a} {Surface brightness (in mag/arcsec$^2$) at 
the effective radius, derived from the best fit model.}
\tablenotetext{b} {K-corrected absolute magnitudes of the nuclei and 
the host galaxies are reported in the $R$-band; no correction for 
galactic extinction is applied.}
\tablenotetext{c} {The N/H ratio refers to the absolute $R$ magnitudes.}
\tablenotetext{d} {The total $U$-band absolute magnitudes were calculated from 
the apparent V magnitudes in Veron-Cetty \& Veron (2003) into our adopted cosmology 
and K-correction. The Galactic extinction was evaluated following 
Schlegel, Finkbeiner \& Davis (1998).} 
\end{deluxetable}

%  TABLE 4
% Average nuc and host luminosities of the samples
\begin{deluxetable}{l l l l l l r l}
\tablecolumns{8}
\tablewidth{0pc}
\tablecaption {TABLE 4}
\tablecaption  {Average properties of the quasar samples.}
\tablehead{
\colhead{sample} & \colhead{N} & \colhead{$<$z$>$}   & 
\colhead{$M(U)_{nucl}$\tablenotemark{a}} & \colhead{$M(R)_{nucl}$} & 
\colhead{$M(R)_{host}$} & \colhead{$(U-R)_{nucl}$} & \colhead{ref\tablenotemark{b}} \\
}
\startdata
HL RQQ &  7 & 1.52$\pm$0.16 & -26.9$\pm$0.9 & -25.9$\pm$1.0 & -23.3$\pm$0.6 & 0.80$\pm$0.24 & F04 \\
HL RLQ & 10 & 1.51$\pm$0.16 & -27.0$\pm$0.7 & -26.6$\pm$0.9 & -24.0$\pm$0.6 & 0.37$\pm$0.88 & F04 \\
\vspace{-0.1cm}\\
LL RQQ &  5\tablenotemark{c} & 1.55$\pm$0.30 & -25.1$\pm$0.6 & -24.1$\pm$0.6 & -22.8$\pm$0.6 & 0.81$\pm$0.86 & this work \\
LL RLQ &  9 & 1.49$\pm$0.25 & -25.2$\pm$0.8 & -25.2$\pm$0.7 & -23.7$\pm$0.8 &-0.16$\pm$0.81 & this work \\
\vspace{-0.1cm}\\
RQQ &  5 & 1.86$\pm$0.12 & -26.2$\pm$0.2 & -24.3$\pm$0.5 & -23.3$\pm$0.9 & 1.94$\pm$0.49 & K01\\
RLQ &  4 & 1.84$\pm$0.14 & -26.6$\pm$0.2 & -25.6$\pm$0.5 & -24.6$\pm$1.4 & 0.93$\pm$0.49 & K01\\
\vspace{-0.1cm}\\
RQQ &  3 & 1.81$\pm$0.07 & -24.0$\pm$0.3 & -23.3$\pm$1.2 & -22.1$\pm$0.7 & 0.68$\pm$1.04 & R01\\
\enddata
\tablenotetext{a} {
The calculated U-band luminosities have been 
K-corrected using the composite quasar SED, under the assumption that the 
underlying host galaxies make negligible contribution to the observed V-band 
magnitudes.
}
\tablenotetext{b} {
References: F04 = Falomo et al. (2004); K01 = Kukula et al. (2001); R01 = Ridgway et al. (2001). 
}
\tablenotetext{c} {
The RQQ TOL 1033.1-27.3 (see the caption of Fig. 3) was excluded from the average.}
\end{deluxetable}

%  TABLE 5
% SPEARMAN RANKCORRELATION COEFFICIENTS
\begin{deluxetable}{l c c c c c c c c c}
\tablecolumns{10}
\tablewidth{0pc}
\tablecaption {TABLE 5}
\tablecaption  {
The Spearman rank correlation coefficients for $M_R$(nucl) vs. $M_R$(host) 
in the quasar samples.
}
\tablehead{
\colhead{} & \colhead{} &\colhead{RQQ} & \colhead{} &\colhead{}&\colhead{RLQ} & \colhead{} &\colhead{(RQQ} &\colhead{+}&\colhead{RLQ)}\\
\colhead{} & \colhead{N\tablenotemark{a}} &\colhead{R$_S$} & \colhead{P(nc)\tablenotemark{b}} &\colhead{N\tablenotemark{a}}&\colhead{R$_S$} & \colhead{P(nc)\tablenotemark{b}} &\colhead{N\tablenotemark{a}} &\colhead{R$_S$}&\colhead{P(nc)\tablenotemark{b}}\\
}
\startdata
HL + LL&12&0.33&0.3 &19&0.38&0.1 &31&0.44&$10^{-2}$\\ %P(nc)= .29, .11, .01
\vspace{-0.2cm}\\
All samples &20&0.25&0.3 &23&0.36&0.1 &43&0.49&$10^{-3}$\\ %P(nc)= .29, .09, .0006
\enddata
\tablenotetext{a} {Number of objects in the considered subsample.}
\tablenotetext{b} {Probability of no correlation, i.e. probability of obtaining the observed R$_S$ values if no correlation is present.}
\end{deluxetable}


\begin{thebibliography}{}

\bibitem[Bahcall et al. 1997]{B97} Bahcall, J.N., Kirhakos S., Saxe D.H., Schneider D.P. 1997, ApJ 479, 642
\bibitem[Barger et al. 2001]{barger01} Barger,A.J., Cowie,L.L., Bautz,M.W., et al., 2001, AJ 122, 2177
\bibitem[]{} Barkhouse,W.A., Hall,P.B., 2001, AJ 121, 2843
\bibitem[Barth 2004]{barth04} Barth,A.J., 2004, The Interplay among Black Holes, Stars and ISM in Galactic Nuclei, Proc. IAU Symp. No. 222 (Ed. T. Storchi-Bergmann, L.C. Ho, and H.R. Schmitt), Cambridge University Press, p.3
\bibitem [Bettoni et al. 2003] {bettoni03} Bettoni,D., Falomo,R., Fasano,G., Govoni,F., 2003, A\&A 399, 869
\bibitem []{} Bower,R.G., Benson,A.J., Malbon,R., et al., 2006, MNRAS 370, 645
\bibitem [Boyle 2001]{boyle01} Boyle, B.J. 2001, Advanced Lectures on the Starburst-AGN Connection, (ed. I.Aretxaga, D.Kunth, R.Mujica), Singapore: World Scientific, p.325
\bibitem  [Bressan et al. 1994]{bressan94} Bressan, A., Chiosi, C., Fagotto, F., 1994, ApJS, 94, 63.
\bibitem [Chary \& Elbaz 2001]{chary01} Chary,R., Elbaz,D., 2001, ApJ 556, 562
\bibitem [Cuby et al. 2000]{cuby} Cuby, J.G., Lidman, C., Moutou, C., Petr, M. 2000, Proc. SPIE, 4008, 1036
\bibitem [Devillard 1999]{devillard99} Devillard, N., 1999, Astronomical Data Analysis Software and Systems VIII, ASP Conference Series, Vol. 172 (ed. D.M. Mehringer, R.L. Plante, D.A. Roberts), p. 333
\bibitem [Dunlop \&  Peacock 1990]{dunlop90}  Dunlop, J. S., Peacock, J. A. 1990 MNRAS, 247, 19
\bibitem [Dunlop et al. 2003]{dunlop03} Dunlop, J.S., McLure, R.J., Kukula, M.J., et al., 2003, MNRAS, 340, 1095
\bibitem [Elvis et al. 1994]{elvis94} Elvis, M., Wilkes, B.J., McDowell, J.C., et al., 1994, ApJS, 95, 1
\bibitem [Falomo et al. 2003]{falomo03} Falomo, R. Carangelo,N., Treves, A. 2003, MNRAS 343, 505
\bibitem [Falomo et al. 2004]{falomo04} Falomo, R. Kotilainen, J.K., Pagani, C., Scarpa, R., Treves, A. 2004, ApJ 604, 495 
\bibitem [Falomo et al. 2005]{falomo05} Falomo, R. Kotilainen, J.K., Scarpa, R., Treves, A. 2005, A\&A 434, 469
\bibitem [Falomo et al. 2007]{falomo07} Falomo, R. Kotilainen, J.K., Scarpa, R., Treves, A. 2007, A\&A, submitted
\bibitem [Ferrarese \& Merritt 2000]{ferrarese00} Ferrarese,L., Merritt,D., 2000, ApJ 539, L9
\bibitem [Ferrarese 2006]{ferrarese06} Ferrarese,L., 2006, Joint Evolution of Black Holes and Galaxies, eds. M. Colpi et al. (Taylor \& Francis Group), p. 1
\bibitem [Floyd et al. 2004] {floyd04} Floyd,D.J.E., Kukula,M.J., Dunlop,J.S., et al., 2004, MNRAS 355, 196
\bibitem [Franceschini et al. 1999] {franc99} Franceschini, A., Hasinger G., Miyaji T., Malquori D. 1999, MNRAS, 310, L5
\bibitem []{} Francis, P.J., Hewett, P.C., Foltz, C.B., et al., 1991, ApJ, 373, 465
\bibitem []{} Francis,P.J., Whiting,M.T., Webster,R.L., 2000, PASA 17, 56
\bibitem []{} Gardner,J.P., Sharples,R.M., Frenk,C.S., Carrasco,B.E., 1997, ApJ, 480, L99
\bibitem [Gebhardt et al. 2000]{gebhardt00} Gebhardt,K., Bender,R., Bower,G., et al., 2000, ApJ 539, L13
\bibitem []{} Granato,G.L., De Zotti,G., Silva,L., Bressan,A., Danese,L., 2004, ApJ 600, 580
\bibitem [Hamilton et al. 2002]{hamil02} Hamilton,T.S., Casertano,S., Turnshek,D.A., 2002, ApJ 576, 61
\bibitem [Hooper et al. 1997] {hooper97} Hooper, E.J., Impey C.D. \& Foltz C.B., 1997, ApJ, 480, L95
\bibitem [Hunt et al. 1997] {hunt97} Hunt, L.K., Malkan,M.A., Salvati,M., et al., 1997, ApJS, 108, 229
%\bibitem [Hyv07]{hyvonen07} Hyv\"onen, T., Kotilainen, J.K., \"Orndahl, E., Falomo, R., Uslenghi,M., 2007, A\&A, in press (astro-ph/0610142)
\bibitem [Haring \& Rix 2004]{haring04} H\"aring,N., Rix,H.W., 2004, ApJ 604, L89
\bibitem [Kauffmann \& Haehnelt 2000]{kauffmann00} Kauffmann, G., Haehnelt, M., 2000, MNRAS, 311, 576
\bibitem [Kauffmann et al. 2003]{kauffmann03} Kauffmann,G., Heckman,T.M., Tremonti,C., et al. 2003, MNRAS 346, 1055
\bibitem [Kormendy \& Richstone 1995]{kor95} Kormendy, J., Richstone, D., 1995, ARA\&A, 33, 581
\bibitem [Kotilainen \& Ward 1994]{koti94}Kotilainen, J.K., Ward, M.J., 1994, MNRAS, 266, 953
%\bibitem [Kotilainen et al. 1998]{kfs98}Kotilainen, J.K., Falomo, R., Scarpa, R. 1998,  A\&A, 332, 503 
%\bibitem [Kotilainen \& Falomo 2000]{kotifal00} Kotilainen, J.K., Falomo, R. 2000,  A\&A, 364, 70
\bibitem [Kukula et al. 2001]{kukula01} Kukula, M.J., Dunlop, J.S., McLure, R.J., et al., 2001, MNRAS, 326, 1533
\bibitem [Madau et al. 1998]{madau98} Madau, P., Pozzetti, L., Dickinson, M. 1998 ApJ, 498, 106 
\bibitem [Magorrian et al. 1998]{mago98} Magorrian, J.,Tremaine, S., Richstone, D., et al., 1998, AJ, 115, 2285
\bibitem []{}Mannucci, F., Basile, F., Poggianti, B.M., et al. 2001, MNRAS, 326, 745
\bibitem [Marconi \& Hunt 2003]{marconi03} Marconi,A., Hunt,L.K. 2003, ApJ 589, L21
\bibitem [Marconi et al. 2004]{marconi04} Marconi,A., Risaliti,G., Gilli,R., et al., 2004, MNRAS 351, 169
\bibitem [McLeod \& Rieke 1994]{mcleod94} McLeod, K.K., Rieke, G.H., 1994, ApJ, 431, 137
\bibitem [McLure \& Dunlop 2002]{mclure02} McLure,R.J., Dunlop,J.S., 2002, MNRAS 331, 795 
\bibitem []{} Nakamura,O., Fukugita,M., Yasuda,N., et al., 2003, AJ 125, 1682
\bibitem [Pagani et al. 2003]{pagani03} Pagani, C., Falomo, R., Treves, A., 2003, ApJ 596, 830
\bibitem [Percival et al. 2000]{perci00} Percival, W.J., Miller, L., McLure, R.J., Dunlop, J.S. 2000, MNRAS 322, 843
\bibitem [Ridgway et al. 2001]{ridgway01} Ridgway, S., Heckman, T., Calzetti, D., Lehnert, M. 2001, ApJ, 550, 122
\bibitem [Schlegel et al. 1998]{schlege98}Schlegel D.J., Finkbeiner D.P., Davis M., 1998, ApJ, 500, 525
\bibitem [Steidel et al. 1999]{steidel99} Steidel, C.C., Adelberger, K.L., Giavalisco, M., Dickinson, M., Pettini, M., 1999, ApJ 519, 1
\bibitem [Taylor et al. 1996]{taylor96} Taylor, G.L., Dunlop, J.S., Hughes, D.H., Robson, E.I. 1996, MNRAS, 283, 930
\bibitem [Veron-Cetty \& Veron 2003]{veron03} Veron-Cetty, M.P., Veron, P. 2003, A\&A 412, 399
\bibitem [Warren et al. 1994]{warren94} Warren, S.J., Hewett, P.C., Osmer, P.S. 1994, ApJ, 421, 412
\bibitem [Yu \& Tremaine 2002] {yu02} Yu,Q., Tremaine,S., 2002, MNRAS 335, 965
\end{thebibliography}
\end{document}